\documentstyle[12pt]{article}
\newcommand{\be}{\begin{equation}}
\newcommand{\ee}{\end{equation}}
\newcommand{\bea}{\begin{eqnarray}}
\newcommand{\eea}{\end{eqnarray}}
\newcommand{\non}{\nonumber}
\begin{document}

\begin{titlepage}
\thispagestyle{empty}

\begin{flushright}
ITP-UH -13-97\\
hep-th/9704002
\end{flushright}
\vspace{5mm}

\begin{center}
{\large\bf The Off-Shell Massive Hypermultiplets Revisited}
\end{center}
\vspace{5mm}

\begin{center}
Sergei M. Kuzenko\footnote{Alexander von Humboldt Research Fellow.
On leave from Department of Quantum Field Theory, Tomsk State University,
Tomsk 634050, Russia}
\end{center}

\begin{itemize}

\item[{}] \footnotesize{{\it Institut f\"ur Theoretische Physik,
Universit\"at Hannover,
Appelstr. 2, 30167 Hannover, Germany}}

\end{itemize}

\vspace{2cm}
\begin{abstract}
On the base of the notion of $N=2$ abelian Yang-Mills supergeometry with
constant curvature, we propose an off-shell formulation for
the massive $q$-hypermultiplet and complex $\omega$-hypermultiplet 
in the standard harmonic superspase without central charge variables. 
The corresponding superpropagators are derived.
\end{abstract}
\vfill

\end{titlepage}

\newpage
\setcounter{page}{1}
\noindent
The only known off-shell formulation for the Fayet-Sohnius
massless hypermultiplet \cite{fss,s} is that invented
by GIKOS \cite{gikos} in
$N=2$ harmonic superspace in terms of a single unconstrained
analytic superfield $q^+$ and its conjugate. It was then extended
to the massive case \cite{gikos,gio,ohta} \`{a} la Sohnius \cite{s}
by building into the matter superfields special dependence on extra
bosonic coordinate(s) used to realize the $N=2$ supersymmetry
with central charge(s). In the present letter we describe
another off-shell formulation for the massive hypermultiplet
in which the mass of the hypermultiplet is generated by its
coupling to a constant $N=2$ abelian gauge superfield which could be 
associated
with the spontaneous breakdown of the $U(1)$ factor ($R$-symmetry)
in the automorphism group $U(2)$ of $N=2$ supersymmetry algebra.
This formulation, being equivalent to that given in Refs.
\cite{gikos,gio,ohta}, seems to be more appropriate for developing
manifestly $N=2$ supersymmetric quantum field theory
(see also Ref. \cite{bbiko}). Besides
evident advantages of such a formulation, it allows us to
generate the hypermultiplet mass without passing to higher
dimensions thus staying only in four-dimensional space-time.
In the framework of our formulation, it is a trivial exercise to obtain
{\it correct superpropagators} for the massive $q$-hypermultiplet
and complex $\omega$-hypermultiplet.
These superpropagators can be used to provide an alternative derivation
of the holomorphic effecive potential of the $N=2$ Maxwell multiplet,
which has been recently computed in terms of $N=2$ superfields \cite{bbiko}.

To describe the $N=2$ massless vector multiplet in 
the standard $N=2$ superspace
parametrized by $z^M=(x^m,\theta_i^\alpha,
\bar \theta^i_{\dot{\alpha}})$, one introduces $U(1)$
gauge-covariant derivatives
\be
{\cal D}_M \equiv({\cal D}_m,{\cal D}^i_\alpha,
\bar {\cal D}^{\dot{\alpha}}_i)= D_M + {\rm i}
A_M(z) {\bf e}
\label{1}
\ee
which are subject to the (anti)commutation relations \cite{gsw,s2}
\begin{eqnarray}
&\{{\cal D}^i_\alpha , \bar {\cal D}_{\dot{\alpha}j} \} =
-2{\rm i}\delta^i_j{\cal D}_{\alpha\dot{\alpha}} \nonumber \\
& \{{\cal D}^i_\alpha ,{\cal D}^j_\beta \} =
2{\rm i}\varepsilon_{\alpha\beta}\varepsilon^{ij} \bar W {\bf e} \qquad
\{\bar {\cal D}_{\dot{\alpha}i} ,\bar {\cal D}_{\dot{\beta}j}\}
= 2{\rm i}\varepsilon_{\dot{\alpha}\dot{\beta}}\varepsilon_{ij}
W{\bf e}
\label{2} \\
&\left [ {\cal D}_{\alpha\dot{\alpha}}, {\cal D}^i_{\beta} \right ] =
\varepsilon_{\alpha\beta}(\bar D^i_{\dot{\alpha}}
\bar W) {\bf e}  \qquad
\left [ {\cal D}_{\alpha \dot{\alpha}} ,\bar {\cal D}_{\dot{\beta}i}
\right ] = \varepsilon_{\dot{\alpha}\dot{\beta}} (D_{\alpha i}
W) {\bf e}\nonumber
\end{eqnarray}
and change in the manner
\be
{\cal
D}_M\rightarrow{\cal D}'_M=e^{{\rm i}\tau {\bf e}} {\cal D}_M
e^{-{\rm i}
\tau {\bf e}} \qquad \varphi \rightarrow \varphi'= e^{{\rm i}\tau{\bf e}}
\varphi
\label{3}
\ee
under local $U(1)$ transformations with unconstrained
real parameter $\tau(z)$. Here $\bf e$ denotes the generator of
$U(1)$ and $\varphi(z)$ some $U(1)$-charged matter multiplet.
The chiral superfield strength
$W, \bar D^i_{\dot {\alpha}}W =0$,
is chargeless.

We are interested in a supergeometry of constant curvature
\be
W_0= const\;.
\label{4}
\ee
Such a supergeometry may originate in $N=2$ super
Yang-Mills theories upon the spontaneous breakdown of gauge symmetry.
As is well known, the potential of the gauge multiplet scalar fields
can vanish at some non-zero values of scalars.
(see, for instance, \cite{agh}). If the gauge group is spontaneously
broken to $U(1)$, we then effectively have a $U(1)$ gauge model
with non-zero vacuum expectation for the gauge multiplet scalar.
In $N=2$ superspace language, the classical vacuum state is described
just by the condition $W_0=const$.

In the case of constant curvature (\ref{4}), the gauge freedom can be
used to impose the following gauge conditions \cite{gsw}
\be
{\bf D}^i_\alpha = D^i_\alpha +
{\rm i}\theta^i_\alpha \bar W_0 {\bf e}\qquad
\bar {\bf D}_{\dot{\alpha} i} = {\bar D}_{\dot{\alpha} i}-
{\rm i} \bar \theta_{\dot{\alpha}i} W_0 {\bf e}\qquad
{\bf D}_{\alpha \dot{\alpha}} =
\partial_{\alpha \dot{\alpha}}\;.
\label{5}
\ee
This gauge choice 
can be treated to be super Poincar\'e covariant
provided every supersymmetry transformation
\be
\delta \varphi = {\rm i}(\epsilon^\alpha_i Q^i_\alpha +
\bar \epsilon_{\dot{\alpha}}^i \bar Q^{\dot{\alpha}}_i)\varphi
\label{6}
\ee
is accompanied by special $\epsilon$-dependent gauge
trasformation
\be
\delta \varphi = {\rm i} \tau {\bf e}\varphi \qquad
\tau = \epsilon^\alpha_i \theta^i_\alpha \bar W_0 -
\bar \epsilon_{\dot{\alpha}}^i {\bar \theta}^{\dot{\alpha}}_i W_0\;.
\label{7}
\ee
As a result, the supersymmetry transformation law of the charged
superfield takes the form
\be
\delta \varphi = {\rm i}(\epsilon {\bf Q} +
\bar \epsilon \bar {\bf Q})\varphi
\label{8}
\ee
where
\be
{\bf Q}^i_\alpha = {\rm i}\frac{\partial}{\partial
\theta^\alpha_i}+ \bar \theta^{\dot{\alpha}i}
\partial_{\dot{\alpha}\alpha}+ \theta^i_\alpha \bar W_0 {\bf e}
\qquad
\bar {\bf Q}_{\dot{\alpha} i} = -{\rm i}\frac{\partial}
{\partial \bar \theta^{\dot{\alpha}i}} -  \theta^{\alpha}_i
\partial_{\dot{\alpha}\alpha}- \bar \theta_{\dot{\alpha}i} W_0{\bf e}\;.
\label{9}
\ee
The modified generators represent the $N=2$ supersymmetry
algebra with central charges
\be
\{ {\bf Q}^i_\alpha ,{\bf Q}^j_\beta\} =
2{\rm i}\varepsilon_{\alpha\beta}\varepsilon_{ij} \bar W_0{\bf e}\qquad
\{ \bar {\bf Q}_{\dot{\alpha}i}
\bar {\bf Q}_{\dot{\beta}j} \} =
2{\rm i}\varepsilon_{\dot{\alpha}\dot{\beta}}
\varepsilon_{ij} W_0{\bf e}
\qquad
\{ {\bf Q}^{i}_{\alpha} ,\bar {\bf Q}_{\dot{\alpha}j}\} =
-2{\rm i}\delta^i_j \partial_{\alpha\dot{\alpha}}
\label{10}
\ee
and eq. (\ref{5}) constitutes the corresponding covariant derivatives,
$\{{\bf Q}^i_\alpha , {\bf D}^j_\beta \} =
\{{\bf Q}^i_\alpha , \bar {\bf D}_{\dot{\beta}j}\}
=0$.

Thus, we see that $N=2$ supersymmetric 
matter theory coupled to the external abelian
gauge superfield with constant strength can be reformulated as
a free matter model possessing the $N=2$ supersymmetry with central charges.
All information about the external superfield
is then encoded in the modified supersymmetry generators
${\bf Q}^i_\alpha ,\bar {\bf Q}_{\dot{\alpha}i}$ (\ref{9})
and the corresponding covariant derivatives ${\bf D}^i_\alpha ,
\bar {\bf D}_{\dot{\alpha}i}$ (\ref{5}).

Now, the on-shell massive hypermultiplet can be realized in
terms of a superfield $\varphi^i(z)$ and its conjugate
$\bar \varphi_i(z)$, with $U(1)$ charges $e$ and $(-e)$
respectively, under the constraints
\be
{\bf D}_\alpha^{(i}\varphi^{j)}=
 \bar {\bf D}_{\dot{\alpha}}^{(i}\varphi^{j)}=0\;.
\label{11}
\ee
From here it follows
\be
(\Box + m^2)\varphi^i = 0 \qquad m^2=e^2\bar W_0 W_0
\label{12}
\ee
and the proof is a copy of that given by Sohnius \cite{s}.  
To extend this description off the mass shell, one is to pass
to the harmonic superspace \cite{gikos}.

In the harmonic superspace approach \cite{gikos}, one introduces into
the game new variables $\|u^\pm_i\| \in SU(2)$ destined
to parametrize two-sphere
$SU(2)/U(1)$. This allows one to partially solve the constraints (\ref{2}) as
follows
\be
{\cal D}^+_\alpha = e^{-{\rm i}\Omega{\bf e}} D^+_\alpha
e^{{\rm i} \Omega{\bf e}} \qquad
\bar {\cal D}^+_{\dot{\alpha}} = e^{-{\rm i}\Omega{\bf e}}
\bar D^+_{\dot{\alpha}}{\rm e}^{{\rm i}\Omega{\bf e}}
\label{13}
\ee
where ${\cal D}^\pm_\alpha =
u^\pm_i{\cal D}^i_\alpha$, $\bar {\cal D}^\pm_{\dot{\alpha}} =
u^\pm_i\bar {\cal D}^i_{\dot{\alpha}}$
and similarly for $D^\pm_\alpha$, $\bar D^\pm_{\dot{\alpha}}$.
Here the superfield $\Omega(z,u)$ is real with respect to the 
analyticity-preserving conjugation $\; \breve{} \;\equiv
\,\stackrel{\star}{\bar{}}$ \cite{gikos}, $\breve{\Omega} = \Omega$,
and has vanishing harmonic $U(1)$-charge, 
$\Omega(z,e^{{\rm i}\alpha} u^+,e^{-{\rm i}\alpha} u^-) = \Omega(z, u^+,u^-)$. 
Now, new superfield types can be
introduced, that is covariantly analytic superfields constrained by
\be
{\cal D}^+_\alpha \Phi^{(p)} = {\bar{\cal D}}^+_{\dot{\alpha}}\Phi^{(p)}=0 
\quad \leftrightarrow \quad
\Phi^{(p)}= e^{-{\rm i}\Omega{\bf e}}\phi^{(p)} \qquad
D^{+}_{\alpha} \phi^{(p)} = \bar D^{+}_{\dot{\alpha}}\phi^{(p)}=0\;.
\label{14}
\ee
Here $\Phi^{(p)}$ carries harmonic $U(1)$-charge $p$,
$\Phi^{(p)}(z,e^{{\rm i}\alpha} u^+,e^{-{\rm i}\alpha} u^-) 
=p \Phi^{(p)}(z,u^+,u^-)$. The $\phi^{(p)}$ is 
an unconstrained superfield over
an analytic subspace of the harmonic superspace \cite{gikos} parametrized by
$\zeta_A\equiv\{x^m_A,\theta^{+\alpha},{\bar\theta}^+_{\dot\alpha}\}$
and $u^\pm_i$, where $x^m_A = x^m - 2{\rm i}\theta^{(i}\sigma^m
{\bar \theta}^{j)}u^+_iu^-_j$ and
$\theta^\pm_\alpha=u^\pm_i\theta^i_\alpha$, ${\bar\theta}^\pm_{\dot\alpha}
=u^\pm_i{\bar \theta}^i_{\dot\alpha}$.
In accordance with eqs. (\ref{3}), (\ref{13}), $\Omega$ transforms by  the rule
\be
\Omega'=\Omega + \lambda - \tau,
\label{15}
\ee
with $\lambda$ an unconstrained analytic real superfield,
$D^{+}_{\alpha} \lambda = \bar D^{+}_{\dot{\alpha}}\lambda=0$,
$\breve{\lambda} = \lambda$. Hence
$\phi^{(p)}$ possesses the transformation law
\be
\phi'^{(p)}= e^{{\rm i}\lambda{\bf e}}\phi^{(p)}\;.
\label{16}
\ee
The $\Phi^{(p)}$ and $\phi^{(p)}$ are said to correspond to the $\tau$- and
$\lambda$-frames respectively.

In the $\lambda$-frame, the harmonic derivatives
$D^{\pm\pm}=u^{\pm i} \, \partial / \partial u^{\mp i} $
turn into the $\lambda$-covariant ones
\be
{\cal D}^{\pm\pm}=e^{{\rm i} \Omega{\bf e}} D^{\pm\pm}
e^{-{\rm i} \Omega{\bf e}} \equiv
D^{\pm\pm}+{\rm i} \Gamma^{\pm\pm}{\bf e}
\label{17}
\ee
and the $N=2$ superspace gauge-covariant derivatives ${\cal D}_M$ turn
into $\stackrel{\sim}{{\cal D}}_M = e^{{\rm i} \Omega{\bf e}}
{{\cal D}}_M e^{-{\rm i} \Omega{\bf e}}$, in particular
$\stackrel{\sim}{{\cal D}}{}^+_\alpha = D^+_\alpha$,
$\stackrel{\sim}{{\bar {\cal D}}}{}^+_{\dot \alpha} ={\bar D}^+_{\dot \alpha}$.
The analytic real connection $\Gamma^{++}$
($D^+_\alpha \Gamma^{++}=\bar D^+_{\dot{\alpha}} \Gamma^{++}=0$, 
$\breve{\Gamma}^{++}=\Gamma^{++}$)
appears to be
the single unconstrained prepotential of the theory. All the
rest geometric objects $A_M$ and $\Gamma^{--}$ are 
expressed in terms of $\Gamma^{++}$.

In the case of supergeometry of constant curvature, $\Omega$
and $\Gamma^{\pm \pm}$ read
\be
\Omega_0 = \theta^{-\alpha}\theta^+_\alpha \bar W_0 +
\bar \theta^{-}_{\dot{\alpha}}\bar \theta^{+\dot{\alpha}} W_0
\qquad
\Gamma_0^{\pm \pm}= -\theta^{\pm \alpha}\theta^\pm_\alpha \bar W_0 -
\bar \theta^{\pm}_{\dot{\alpha}}\bar \theta^{\pm \dot{\alpha}} W_0\;.
\label{18}
\ee
We denote by ${\bf D}^{\pm \pm}$ the derivatives (\ref{17})
associated with $\Omega_0$ and $\Gamma_0^{\pm \pm}$.

The Fayet-Sohnius massless hypermultiplet is described in
harmonic superspace by an unconstrained analytic superfields
$q^+(\zeta_A,u)$ \cite{gikos}.
The classical action for the massless hypermultiplet
interacting with the gauge superfield $\Gamma^{++}(\zeta_A, u)$ reads
\be
S=\int {\rm d}\zeta^{(-4)}_A {\rm d}u \breve{q}^+ (D^{++}+ {\rm i}
e\Gamma^{++})q^+ =
\int {\rm d}\zeta^{(-4)}_A {\rm d}u \breve{q}^+ {\cal D}^{++}q^+
\label{19}
\ee
where ${\rm d}\zeta^{(-4)}_A = {\rm d}^{4} x_{A} {\rm d}^{2} \theta^{+}
{\rm d}^2 \bar
\theta^+$.
Because of the identity $({\cal D}^{--})^2{\cal D}^{++}q^+=
{\cal D}^{++}({\cal D}^{--})^2q^+$,
the equation of motion
\be
{\cal D}^{++}q^+ = 0
\label{20}
\ee
implies $({\cal D}^{--})^2q^+ = 0$ 
(since the equation $D^{++}f^{(p)}(u) = 0$ has the unique solution
$f^{(p)} = 0$ for $p < 0$),
and therefore
it leads to the following consequence  
\bea
&&{\cal H}q^{+}=-\frac{1}{32}
\stackrel{\sim}{{\bar {\cal D}}}{}^+_{\dot{\alpha}}
\stackrel{\sim}{{\bar {\cal D}}}{}^{+\dot{\alpha}} 
\stackrel{\sim}{{\cal D}}{}^{+\alpha} 
\stackrel{\sim}{{\cal D}}{}^{+}_{\alpha}
({\cal D}^{--})^2 q^+ = 0
\nonumber
\\
&&{\cal H} = \stackrel{\sim}{{\cal D}}{}^{m}
\stackrel{\sim}{{\cal D}}_{m}
+ \frac{{\rm i}}{2}e(D^{+ \alpha} W)
\stackrel{\sim}{{\cal D}}{}^-_\alpha
+ \frac{{\rm i}}{2}e(\bar
D^{+}_{\dot{\alpha}} \bar W)
\stackrel{\sim}{{\bar {\cal D}}}{}^{-\dot{\alpha}} 
\label{21}
\\
&&- \frac{{\rm i}}{4}e(\bar D^{+}_{\dot{\alpha}}\bar D^{+\dot{\alpha}}
\bar W){\cal D}^{- -} +
\frac{{\rm i}}{4}e(D^{- \alpha}D^+_\alpha W) + e^2\bar W W\;.
\nonumber
\eea
Here all the gauge-covariant derivatives are taken in the $\lambda$-frame.

If we now consider the coupling of $q^+$ to the gauge superfield of constant
curvature (\ref{4}), then the last equation turns, in the supersymmetric 
gauge (\ref{5}, \ref{18}),
into
\be
(\partial^m\partial_m + e^2 \bar W_0 W_0)q^+ =0\;.
\label{22}
\ee
We see that the interaction with constant background 
makes the hypermultiplet massive, and the value of mass coincides with
that of the central charge induced. It is worth noting that the dynamical
equation ${\bf D}^{++} q^+ =0$ implies
\be
q^+(x,\theta,u)=e^{{\rm i} e \Omega_0(\theta,u)} u^+_i\varphi^i(x,\theta)
\label{333}
\ee
with $\Omega_0$ given in eq. (\ref{18}) and $\varphi^i$ being restricted 
by the on-shell constraints (\ref{11}).

It has been just shown how to formulate the free massive Fayet-Sohnius
hypermultiplet. To describe the massive hypermultiplet coupled to
the Yang-Mills superfield \cite{gikos} $V^{++} = V^{++a}(\zeta_A, u)T^a$,
which is associated with some gauge group $G$ (with $T$'s the generators
of $G$ in the matter representation), one is to consider the
massless hypermultiplet coupled to Yang-Mills superfield
$(\Gamma_0^{++}, V^{++})$ for the gauge group $U(1)\times G$, where
$\Gamma^{++}_0$ (\ref{18}) describes the constant curvature supergeometry.
We thus arrive at the action
\be
S=\int {\rm d}\zeta^{(-4)}_A {\rm d}u \breve{q}^+ (D^{++}+ {\rm i}
e\Gamma_0^{++} +{\rm i}V^{++} )q^+
=\int {\rm d}\zeta^{(-4)}_A {\rm d}u \breve{q}^+ ({\bf D}^{++}
+{\rm i}V^{++} )q^+\;.
\label{24}
\ee

There is another, basically equivalent description of the irreducible 
massless hypermultiplet in terms of an unconstrained analytic real 
superfield $\omega(\zeta_A,u)$ \cite{gikos}. 
When $\omega$ is taken to be complex, however, it can be made
massive on the base of 
the same mechanism of generating the mass  
we have just described for the $q$-hypermultiplet.
Let us consider the complex $\omega$-hypermultiplet coupled to 
the abelian gauge superfield $V^{++}$. The gauge invariant action reads
\be
S=
-\int {\rm d}\zeta^{(-4)}_A
{\rm d}u\,{\cal D}^{++}\breve{\omega}{\cal D}^{++}\omega =
-\int {\rm d}\zeta^{(-4)}_A
{\rm d}u\,(D^{++} - {\rm i}e V^{++})\breve{\omega}
(D^{++} + {\rm i}e V^{++})\omega \;.
\label{o}
\ee
Because of the identity $({\cal D}^{--})^2({\cal D}^{++})^2\omega=
({\cal D}^{++})^2({\cal D}^{--})^2\omega$,
the dynamical equation 
\be
({\cal D}^{++})^2 \omega = 0
\label{kuku}
\ee
implies $({\cal D}^{--})^2 \omega = 0$, and therefore
it leads to the following consequence 
\be
{\cal H} \omega =0
\label{oh}
\ee
with ${\cal H}$ the same second-order differential operator as in eq.
(\ref{21}). When $V^{++}=V_0^{++}$, this equation takes the form (\ref{22}).
It is worth pointing out that in this case the dynamical equation 
$({\bf D}^{++})^2 \omega =0$ leads to
\be
\omega(x,\theta,u)=e^{{\rm i} e \Omega_0(\theta,u)} 
\left\{\psi(x,\theta) +\psi^{(ij)}(x,\theta)u^+_iu^-_j\right\}
\label{222}
\ee
with $\psi$ and $\psi^{ij}$ constrained by
\be
{\bf D}^i_\alpha \psi = \frac{1}{3}{\bf D}_{\alpha j} \psi^{ij}
\qquad {\bf D}_\alpha^{(i} \psi^{jk)} = 0
\label{mu}
\ee
as well as by similar constraints but with ${\bf D}$'s replaced by
${\bar {\bf D}}$'s.

Before turning to the structure of superpropagators, let us consider
the $N=2$ supersymmetric massless electrodynamics
with the action
\be
S_{{\rm SED}}=\frac{1}{2}\int {\rm d}^4 x {\rm d}^4 \theta W^2
+ \int {\rm d}\zeta^{(-4)}_A
{\rm d}u \breve{q}^+ (D^{++}+ {\rm i}e\Gamma^{++})q^+\;.
\label{25}
\ee
Because of the Bianchi identity
\be
D^{\alpha (i}D^{j)}_\alpha W = \bar D^{(i}_{\dot{\alpha}}
\bar D^{j)\dot{\alpha}}
\bar W
\label{26}
\ee
the first term in $S_{{\rm SED}}$ remains unchanged under the shift
$W \rightarrow W=W_0+W$, $W_0=const$. As to the second term, it
gets modified by the replacement
\be
\Gamma^{++} \rightarrow \Gamma^{++}_0 + \Gamma^{++}
\label{27}
\ee
with $\Gamma_0^{++}$ as in eq.(\ref{18}).
Such a substitution neither breaks the
gauge invariance nor $N=2$ supersymmetry.  But for $W_0\neq0$ we
have the supersymmetry with central charges in the matter sector,
and hence the matter superfields becomes massive.

The propagators of the massive $q$- and $\omega$-hypermultiplets 
can be readily obtained on the base of
the known results for the free massless
hypermultiplets \cite{gios}. The Feynman propagators read
\bea
G_{\rm F}^{(1,1)}(1,2) &\equiv & {\rm i}<q^+(1)\,\breve{q}^+(2)> \label{qf}  \\
&=& {1\over {\Box_1 +m^2}}(D_1^+)^4(D_2^+)^4 \left\{
\delta^4(x_1-x_2)\delta^8(\theta_1-\theta_2)
\frac{1}{(u^+_1 u^+_2)^3}
e^{{\rm i}e(\Omega_0(1)-\Omega_0(2))} \right\}\non \\
G_{\rm F}^{(0,0)}(1,2) &\equiv& {\rm i}\,<\omega(1)\,\breve{\omega}(2)> 
\label{of} \\
&=& \frac{1}{\Box_1 +m^2}
(D_1^+)^4(D_2^+)^4 \left\{\delta^4(x_1-x_2)\delta^8(\theta_1-\theta_2)
{(u^-_1 u^-_2)\over (u^+_1 u^+_2)^3}
e^{{\rm i}e(\Omega_0 (1)-\Omega_0 (2))}\right\}
\non
\eea
and satisfy the equations
\bea
{\bf D}_1^{++}G_{\rm F}^{(1,1)}(1,2)&=&- \delta_A^{(3,1)}(1,2)
\label{28}\\
({\bf D}_1^{++})^2G_{\rm F}^{(0,0)}(1,2)&=& -\delta_A^{(4,0)}(1,2)
\label{29}
\eea
where $1,2\equiv (\zeta_{1,2A},u_{1,2})$, 
$(D^+)^4= 1/16\, (D^+)^2({\bar D}^+)^2$
\be 
\delta_A^{(n,4-n)}(1,2)= 
(D_2^+)^4 \left\{
\delta^4(x_1-x_2)\delta^8(\theta_1-\theta_2)
\delta^{(n,-n)}(u_1,u_2)\right\}
\label{del1}
\ee
 and $\delta^{(n,-n)}(u_1,u_2)$
is the proper harmonic $\delta$-function \cite{gios}.
In contrast to the free massless superpropagator \cite{gios},
$G_{\rm F}^{(1,1)}(1,2)$ is no more antisymmetric in its arguments
\be
G_{\rm F}^{(1,1)}(1,2) \neq - G_{\rm F}^{(1,1)}(2,1)\;.
\label{30}
\ee
This is a consequence of the fact that $q^+$ and $\breve{q}^+$ correspond to
different eigenvalues, $e$ and $-e$, of the $U(1)$-generator ${\bf e}$
entering the supersymmetry generators (\ref{9}). The superpropagator
of $q$-hypermultiplet was originally obtained by Zupnik \cite{z}
by means of reduction from six-dimentional space-time.

Let us comment on the derivation of eqs. (\ref{qf}) and (\ref{of})
since there have been given examples of 
quantum calculations with the use of incorrect
massive $q$-superpropagator. The main steps here are in principle
the same as those applied to deduce eq. (\ref{21}) from (\ref{20}).  
We start with the equation on the $q$-superpropagator (\ref{28}) and
act on both sides by $({\bf D}_1^{--})^2$. Because of the identity
$({\bf D}_1^{--})^2{\bf D}_1^{++}G_{\rm F}^{(1,1)}(1,2)=
{\bf D}_1^{++}({\bf D}_1^{--})^2G_{\rm F}^{(1,1)}(1,2)$ and taking into 
account the relation
\be
({\bf D}_1^{--})^2 \delta_A^{(3,1)}(1,2) =
{\bf D}_1^{++}\left\{ 2(D^+_2)^4 \delta^{12}(z_1-z_2)
\frac{1}{(u^+_1 u^+_2)^3}
e^{{\rm i}{\bf e}(\Omega_0(1)-\Omega_0(2))} \right\}
\ee
which follows from the definition ${\bf D}^{\pm\pm}=
e^{{\rm i}{\bf e}\Omega_0} D^{\pm\pm}e^{-{\rm i}{\bf e}\Omega_0}$
and the identity \cite{gios}
\be
(D_1^{--})^2 \delta^{(3,-3)}(u_1,u_2) = 2D_1^{++} \frac{1}{(u^+_1u^+_2)^3}
\ee
we may conclude
\be
({\bf D}_1^{--})^2G_{\rm F}^{(1,1)}(1,2)=
-2(D_2^+)^4 \left\{
\delta^{12}(z_1-z_2)
\frac{1}{(u^+_1 u^+_2)^3}
e^{{\rm i}e(\Omega_0(1)-\Omega_0(2))} \right\}\;. 
\ee
It remains to act on both sides of this relation by $(D_1^+)^4$ and then to
make use of eqs. (\ref{21}), (\ref{4}).

The propagators obtained can be used for calculating the low-energy
effective action in $N=2$ super Yang-Mills theories with spontaneously
broken gauge symmetry in manifestly supersymmetric way. 
This work is in progress.

\vspace{1cm}

\noindent
{\bf Acknowledgements}
I thank I.L. Buchbinder for collaboration at the early stage of this
work and acknowledge fruitful discussions with 
E.I. Buchbinder and N. Dragon. This work was supported by
the RFBR-DFG project No 96-02-00180,
the RFBR project No 96-02-16017 and by the Alexander
von Humboldt Foundation.


\end{document}